\documentclass[twocolumn,twoside]{revtex4}
\usepackage{graphicx}
\usepackage{fancyhdr}
\usepackage{upgreek} 
\usepackage{ifthen} 
\newboolean{uprightparticles}
\setboolean{uprightparticles}{false} 

\usepackage{xspace} 
\usepackage{upgreek}

\def\lhcb {\mbox{LHCb}\xspace}

\def\babar  {\mbox{BaBar}\xspace}
\def\belle  {\mbox{Belle}\xspace}

\def\MagUp {\mbox{\em Mag\kern -0.05em Up}\xspace}

\ifthenelse{\boolean{uprightparticles}}%
{

 \def\Pmu         {\ensuremath{\upmu}\xspace}                 
 \def\Pnu         {\ensuremath{\upnu}\xspace}

 \def\Ptau        {\ensuremath{\uptau}\xspace}

 \def\PDelta      {\ensuremath{\Delta}\xspace}                 
 \def\PXi      {\ensuremath{\Xi}\xspace}                 
 \def\PLambda      {\ensuremath{\Lambda}\xspace}                 
 \def\PSigma      {\ensuremath{\Sigma}\xspace}                 
 \def\POmega      {\ensuremath{\Omega}\xspace}                 
 \def\PUpsilon      {\ensuremath{\Upsilon}\xspace}                 
 
 \def\PB      {\ensuremath{\mathrm{B}}\xspace}                 
                  
 \def\PD      {\ensuremath{\mathrm{D}}\xspace}

 \def\PK      {\ensuremath{\mathrm{K}}\xspace}

 \def\PW      {\ensuremath{\mathrm{W}}\xspace}

 \def\PZ      {\ensuremath{\mathrm{Z}}\xspace}                 
                  
 \def\Pb      {\ensuremath{\mathrm{b}}\xspace}                 
 \def\Pc      {\ensuremath{\mathrm{c}}\xspace}

 \def\Pi      {\ensuremath{\mathrm{i}}\xspace}

}
{

 \def\Pmu         {\ensuremath{\mu}\xspace}                 
 \def\Pnu         {\ensuremath{\nu}\xspace}

 \def\Ptau        {\ensuremath{\tau}\xspace}

 \mathchardef\PDelta="7101
 \mathchardef\PXi="7104
 \mathchardef\PLambda="7103
 \mathchardef\PSigma="7106
 \mathchardef\POmega="710A
 \mathchardef\PUpsilon="7107
                  
 \def\PB      {\ensuremath{B}\xspace}                 
                  
 \def\PD      {\ensuremath{D}\xspace}

 \def\PK      {\ensuremath{K}\xspace}

 \def\PW      {\ensuremath{W}\xspace}

 \def\PZ      {\ensuremath{Z}\xspace}                 
                  
 \def\Pb      {\ensuremath{b}\xspace}                 
 \def\Pc      {\ensuremath{c}\xspace}

 \def\Pi      {\ensuremath{i}\xspace}

}

\DeclareRobustCommand{\optbar}[1]{\shortstack{{\miniscule (\rule[.5ex]{1.25em}{.18mm})}
  \\ [-.7ex] $#1$}}

\def\mun        {{\ensuremath{\Pmu^-}}\xspace} 

\def\taum       {{\ensuremath{\Ptau^-}}\xspace}

\def\ellm       {{\ensuremath{\ell^-}}\xspace}

\def\neu        {{\ensuremath{\Pnu}}\xspace}
\def\neub       {{\ensuremath{\overline{\Pnu}}}\xspace}

\def\neumb      {{\ensuremath{\neub_\mu}}\xspace}

\def\neutb      {{\ensuremath{\neub_\tau}}\xspace}
\def\neul       {{\ensuremath{\neu_\ell}}\xspace}
\def\neulb      {{\ensuremath{\neub_\ell}}\xspace}

\def\W      {{\ensuremath{\PW}}\xspace}

\def\Z      {{\ensuremath{\PZ}}\xspace}

\def\cquark    {{\ensuremath{\Pc}}\xspace}

\def\bquark    {{\ensuremath{\Pb}}\xspace}

  \def\Kbar    {{\kern 0.2em\overline{\kern -0.2em \PK}{}}\xspace}

\def\KorKbar    {\kern 0.18em\optbar{\kern -0.18em K}{}\xspace}

  \def\Dbar    {{\kern 0.2em\overline{\kern -0.2em \PD}{}}\xspace}
\def\D       {{\ensuremath{\PD}}\xspace}

\def\DorDbar    {\kern 0.18em\optbar{\kern -0.18em D}{}\xspace}
\def\Dz      {{\ensuremath{\D^0}}\xspace}

\def\Dp      {{\ensuremath{\D^+}}\xspace}

\def\Dstar   {{\ensuremath{\D^*}}\xspace}

\def\B       {{\ensuremath{\PB}}\xspace}
\def\Bbar    {{\ensuremath{\kern 0.18em\overline{\kern -0.18em \PB}{}}}\xspace}

\def\BorBbar    {\kern 0.18em\optbar{\kern -0.18em B}{}\xspace}

\def\Bzb     {{\ensuremath{\Bbar{}^0}}\xspace}

\def\Bub     {{\ensuremath{\B^-}}\xspace}

\def\Bm      {{\ensuremath{\Bub}}\xspace}

  \def\Y#1S{\ensuremath{\PUpsilon{(#1S)}}\xspace}

\def\Lbar        {{\ensuremath{\kern 0.1em\overline{\kern -0.1em\PLambda}}}\xspace}
\def\LorLbar    {\kern 0.18em\optbar{\kern -0.18em \PLambda}{}\xspace}

\def\BF         {{\ensuremath{\mathcal{B}}}\xspace}

\def\BR         {\BF}

\def\to                 {\ensuremath{\rightarrow}\xspace}

\def\qsq       {{\ensuremath{q^2}}\xspace}

\def\AT#1     {\ensuremath{A_{\mathrm{T}}^{#1}}\xspace}           

\def\C#1      {\ensuremath{\mathcal{C}_{#1}}\xspace}                       
\def\Cp#1     {\ensuremath{\mathcal{C}_{#1}^{'}}\xspace}                    
\def\Ceff#1   {\ensuremath{\mathcal{C}_{#1}^{\mathrm{(eff)}}}\xspace}        
\def\Cpeff#1  {\ensuremath{\mathcal{C}_{#1}^{'\mathrm{(eff)}}}\xspace}       
\def\Ope#1    {\ensuremath{\mathcal{O}_{#1}}\xspace}                       
\def\Opep#1   {\ensuremath{\mathcal{O}_{#1}^{'}}\xspace}                    

\newcommand{\tev}{\ifthenelse{\boolean{inbibliography}}{\ensuremath{~T\kern -0.05em eV}}{\ensuremath{\mathrm{\,Te\kern -0.1em V}}}\xspace}
\newcommand{\gev}{\ensuremath{\mathrm{\,Ge\kern -0.1em V}}\xspace}
\newcommand{\mev}{\ensuremath{\mathrm{\,Me\kern -0.1em V}}\xspace}
\newcommand{\kev}{\ensuremath{\mathrm{\,ke\kern -0.1em V}}\xspace}
\newcommand{\ev}{\ensuremath{\mathrm{\,e\kern -0.1em V}}\xspace}
\newcommand{\gevc}{\ensuremath{{\mathrm{\,Ge\kern -0.1em V\!/}c}}\xspace}
\newcommand{\mevc}{\ensuremath{{\mathrm{\,Me\kern -0.1em V\!/}c}}\xspace}
\newcommand{\gevcc}{\ensuremath{{\mathrm{\,Ge\kern -0.1em V\!/}c^2}}\xspace}
\newcommand{\gevgevcccc}{\ensuremath{{\mathrm{\,Ge\kern -0.1em V^2\!/}c^4}}\xspace}
\newcommand{\mevcc}{\ensuremath{{\mathrm{\,Me\kern -0.1em V\!/}c^2}}\xspace}

\newcommand{\stat}{\ensuremath{\mathrm{\,(stat)}}\xspace}
\newcommand{\syst}{\ensuremath{\mathrm{\,(syst)}}\xspace}

\def\gsim{{~\raise.15em\hbox{$>$}\kern-.85em
          \lower.35em\hbox{$\sim$}~}\xspace}
\def\lsim{{~\raise.15em\hbox{$<$}\kern-.85em
          \lower.35em\hbox{$\sim$}~}\xspace}

\def\ptot       {\mbox{$p$}\xspace}

\def\evtgen     {\mbox{\textsc{EvtGen}}\xspace}

\def\photos     {\mbox{\textsc{Photos}}\xspace}

\def\pythia     {\mbox{\textsc{Pythia}}\xspace}

\def\tell1  {TELL1\xspace}
\def\ukl1   {UKL1\xspace}

\def\RDp      {{\ensuremath{\mathcal{R}(\Dp)}}\xspace}
\def\RDz      {{\ensuremath{\mathcal{R}(\Dz)}}\xspace}
\def\RD       {{\ensuremath{\mathcal{R}(\D)}}\xspace}
\def\Hc       {{\ensuremath{H_c}}\xspace}
\def\Hb       {{\ensuremath{H_b}}\xspace}
\def\RHc       {{\ensuremath{\mathcal{R}(\Hc)}}\xspace}
\def\RDst     {{\ensuremath{\mathcal{R}(\Dstar)}}\xspace}

\def\emax     {{\ensuremath{E_{\rm{max}}}}\xspace}
\def\emu      {{\ensuremath{E_{\mu}}}\xspace}
\def\mmiss    {{\ensuremath{m_{\rm miss}^2}}\xspace}
\def\dQED     {{\ensuremath{\delta_{\rm{QED}}}}\xspace}
\def\OmegaC   {{\ensuremath{\Omega_{\rm{C}}}}\xspace}
\def\qsq      {{\ensuremath{q^2}}\xspace}

\graphicspath{{./figs/}} 

\begin{document}

\title{Impacts of radiative corrections on measurements 
\\of lepton flavour universality in $\B \to \D \ell \neul$ decays}

\author{S. Klaver}
\affiliation{INFN Laboratori Nazionali di Frascati, Via Enrico Fermi, 40, 00044 Frascati, Italy}

\begin{abstract}
Radiative corrections to $\B \to \D \ell \neul$ decays can have an impact on predictions and measurements of the lepton universality ratios \RDp and \RDz. 
These proceedings summarise a study on the comparison between recent calculations of soft-photon corrections on these ratios and the corrections simulated
by the \photos package. Also the impact of Coulomb interactions, not simulated in \photos, is discussed. Using pseudo-experiments, the effect of high-energy photon
emission is studied in an \lhcb-like environment, showing a bias of up to 7\% on measurements of \RD.
\end{abstract}

\maketitle

\thispagestyle{fancy}

\section{Introduction}
In the Standard Model (SM) it is assumed that the only difference between the three generations of leptons is their mass, and that their gauge couplings are the 
same. This assumption, called lepton 
universality (LU), can be tested by measuring the ratio of decay rates, which ensures that many experimental and theoretical uncertainties are cancelled in the ratio.
One type of these LU measurements is performed using semileptonic \B decays of the form
$\bquark\to\cquark\ellm\neulb$, commonly known as measurements of \RHc. This is defined as
\begin{equation}
    \RHc = \frac{\BR(\Hb\to\Hc\taum\neutb)}{\BR(\Hb\to\Hc\ell^-\neulb)} \, ,
\end{equation}
where \Hb and \Hc are a \bquark and \cquark hadron, respectively, and $\ell$ is either 
an electron or muon.

Measurements of \RHc have been performed by the \lhcb, \belle and \babar experiments.
For \RD, the average of the measured value of \RD is
$0.349\pm 0.027\stat \pm 0.015\syst$ \cite{Lees:2012xj,Huschle:2015rga,Abdesselam:2019dgh}.
The predicted value for \RD, assuming isospin symmetry, is 
$\RD = \RDp = \RDz = 0.299 \pm 0.003$~\cite{Bigi:2016mdz,Bernlochner:2017jka,Jaiswal:2017rve,Aoki:2019cca}.
Even though \RD differs from the SM prediction by only 1.4$\sigma$, the deviation from the SM of the 
combined \RD and \RDst is about 3.1$\sigma$~\cite{HFLAV16}.

Radiative corrections were long thought to be negligible
at the level of precision of measurements and predictions of \RD.
Recently, however, de Boer et al.~\cite{deBoer:2018ipi} presented a new evaluation of the 
long-distance electromagnetic (QED) contributions
to $\Bzb\to\Dp\ellm\neulb$ and $\Bm\to\Dz\ellm\neulb$ decays, where $\ellm = \mun, \taum$. 
These corrections are different for \mun and \taum decays, such that they do not cancel in \RD.
A proper evaluation of the radiative corrections can alter SM predictions and increase their uncertainty. 

In measurements of \RD, radiative corrections are simulated using the \photos package~\cite{Golonka:2005pn,Golonka:1379813}.
These proceedings, which are a summary of the studies described in Ref.~\cite{Cali:2019nwp}, show the difference between the QED 
corrections simulated in \photos and those predicted by Ref.~\cite{deBoer:2018ipi}.
They describe a study on the effects of under- or overestimating these corrections in simulation on measurements of \RD. 

\section{Radiative corrections in P{\small HOTOS}}
\label{sec:comparePhotos}

\photos~\cite{Golonka:2005pn,Golonka:1379813} is a universal MC algorithm that simulates effects of QED corrections.
The corrections simulated by \photos have successfully been tested for \W, \Z, and \B decays 
and should be tested for every type of measurements individually, especially when high precision is needed. 
Unlike Ref.~\cite{deBoer:2018ipi}, \photos does not include Coulomb corrections. These corrections concern the enhancement of decay rates 
due to the interaction of two charged particles and are therefore relevant for the \Dp decay, but not for the \Dz decay.

Recent versions of \photos include multi-photon emissions as well as interference between final-state photons, whereas Ref.~\cite{deBoer:2018ipi}
also includes the interference between initial- and final state photons. 
The calculation by de Boer et al. in Ref.~\cite{deBoer:2018ipi} is valid in the regime
in which the maximum energy of the radiated photons is smaller than the lepton mass, the muon mass in this case. \photos also includes photon 
emission with higher energies. Neither Ref.~\cite{deBoer:2018ipi} nor \photos include the emission of photons depending on the hadronic structure.
These so called structure-dependent photons impact the spin of the decay particle and may also interfere with bremsstrahlung photons.

\begin{figure*}[t]
    \includegraphics[width=0.32\linewidth]{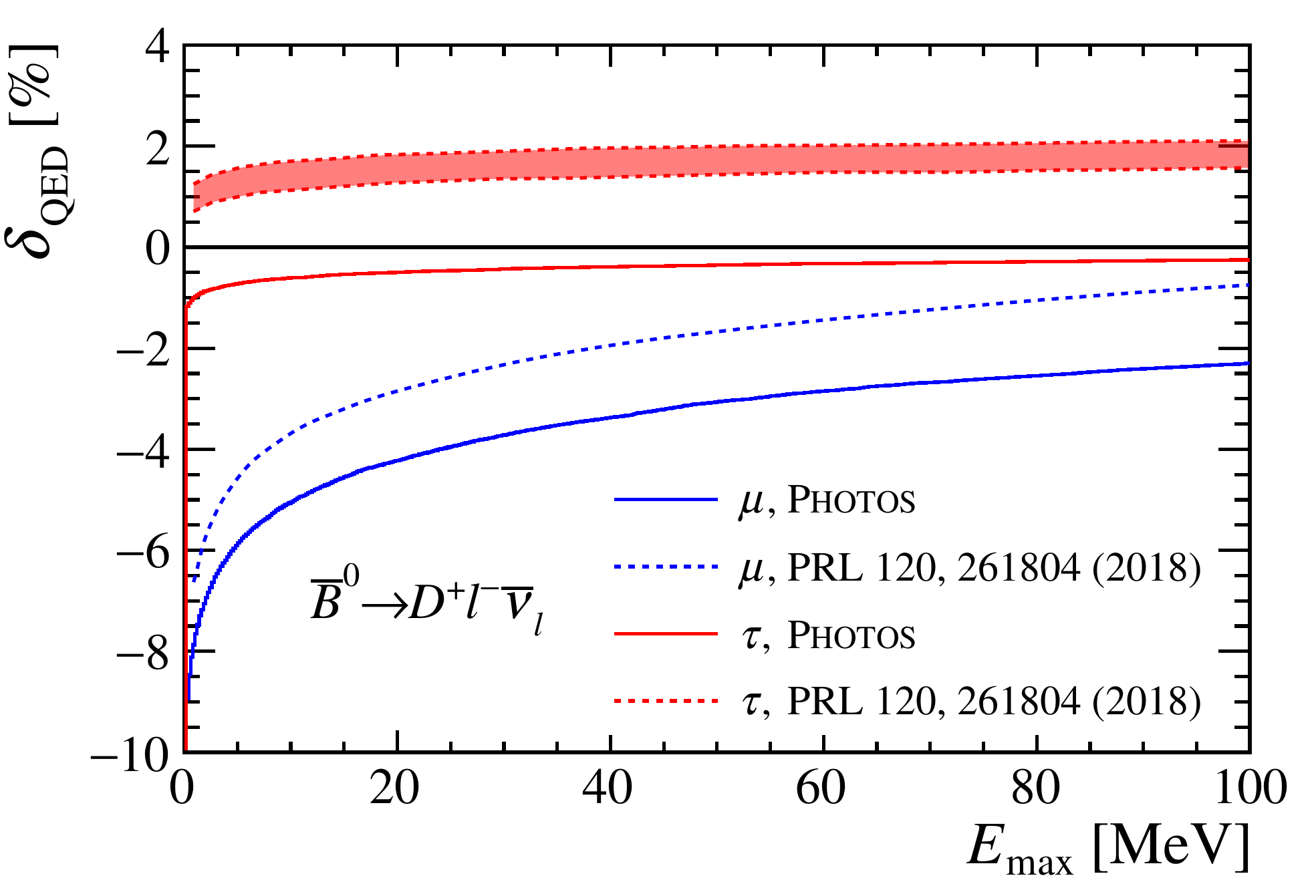} 
    \includegraphics[width=0.32\linewidth]{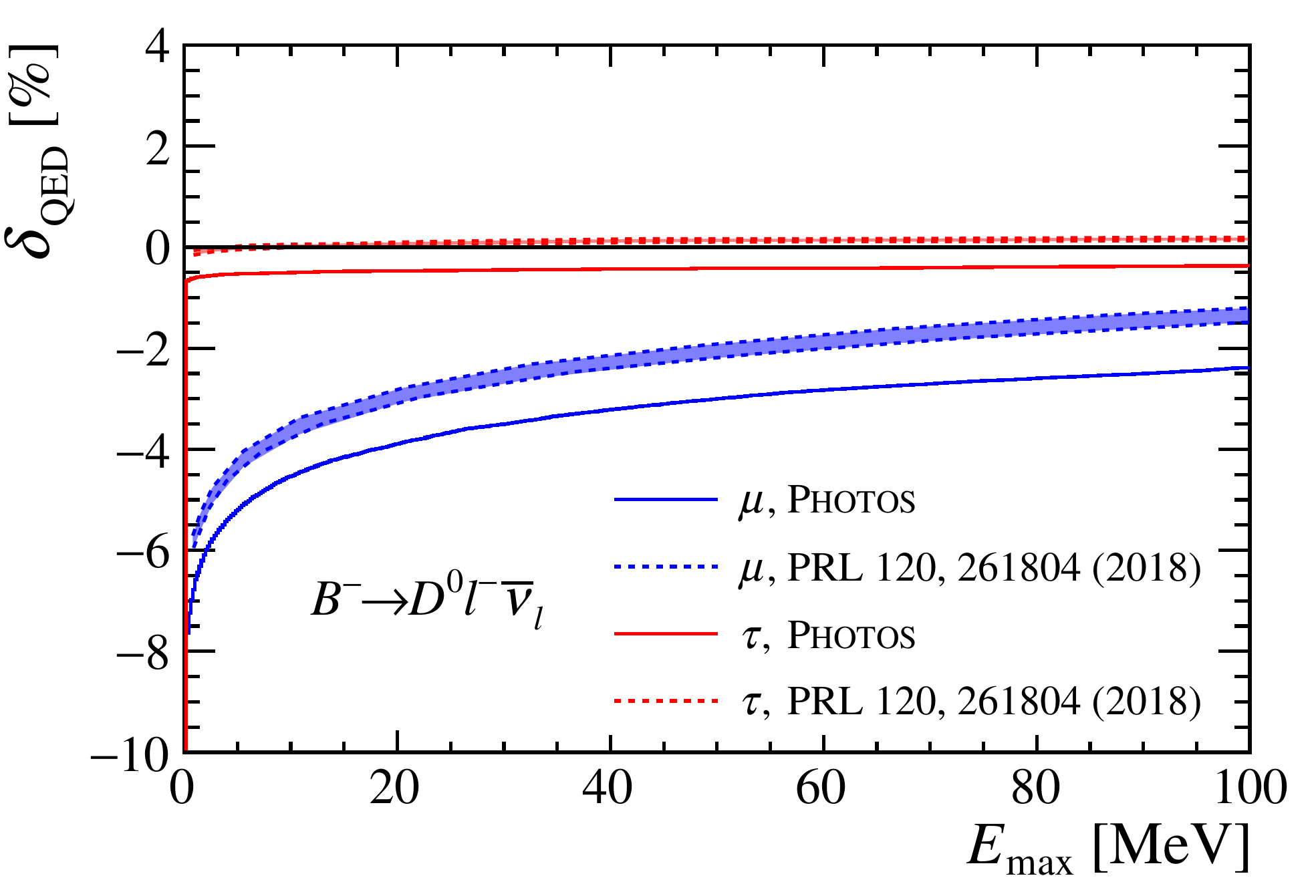} 
    \includegraphics[width=0.32\linewidth]{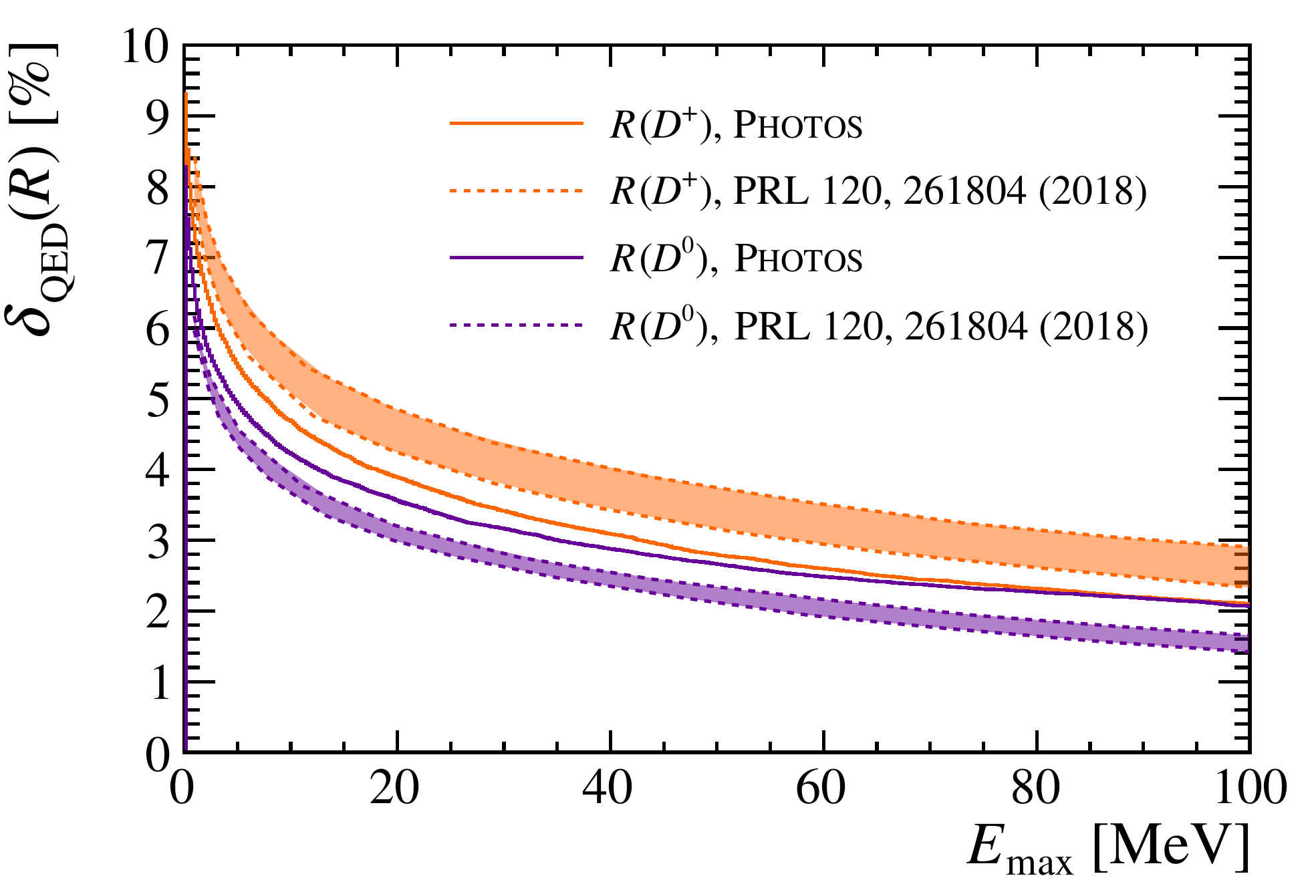}
  \vspace{-0.2cm}
  \caption{
  Radiative corrections to the branching ratios of 
  $\bar{B}^0\to D^+\ell^-\bar{\nu_\ell}$ (left) and $B^-\to D^0\ell^-\bar{\nu_\ell}$ (middle) decays, as a function of \emax. 
  The long-distance QED corrections to \RDp (orange) and \RDz 
  (violet) as a function of \emax (right).
  The plots are obtained from simulated data (solid 
  lines) and from Ref.~\cite{deBoer:2018ipi} (dashed lines).}
  \label{fig:RD0plots}
\end{figure*}

\begin{figure*}
    \includegraphics[width=0.32\linewidth]{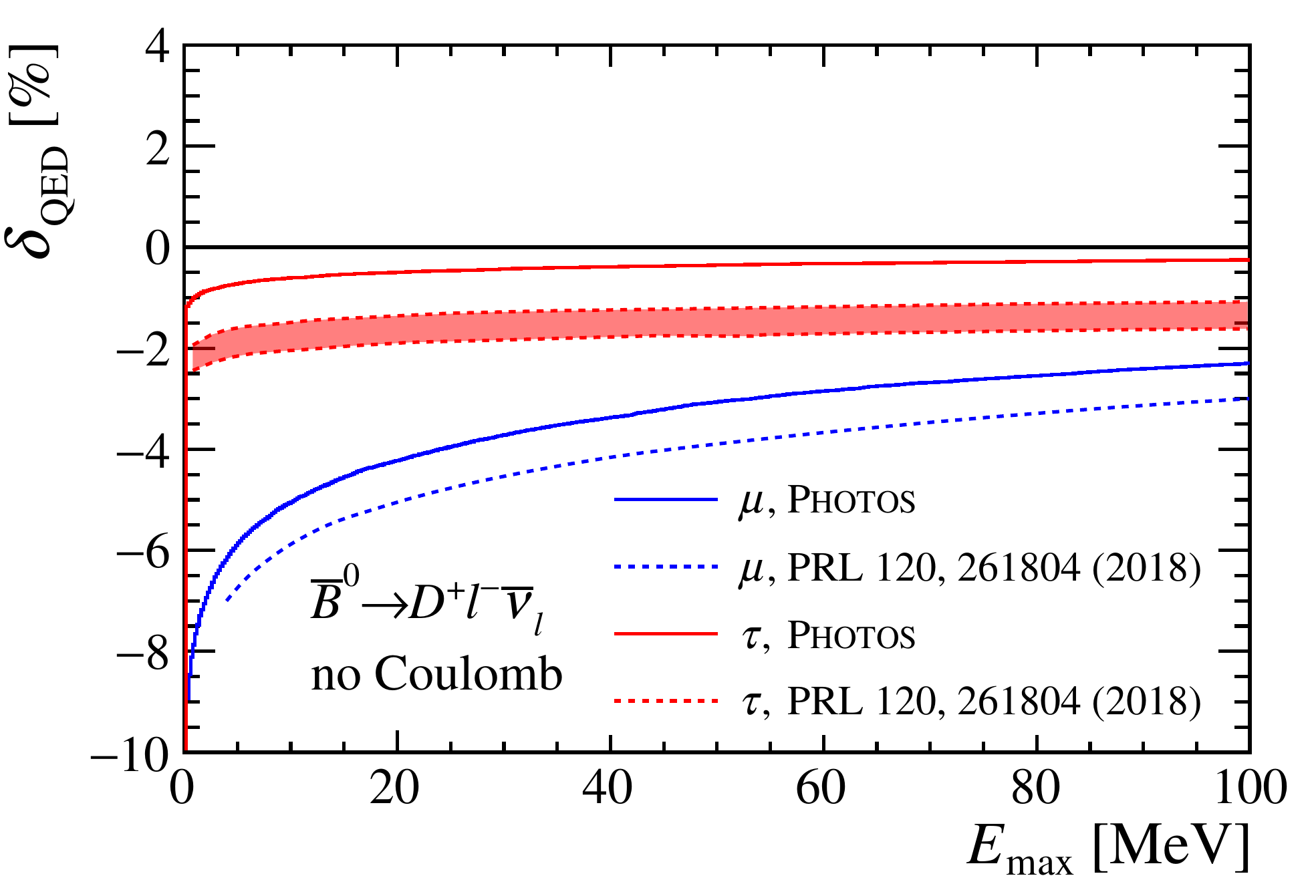}
    \includegraphics[width=0.32\linewidth]{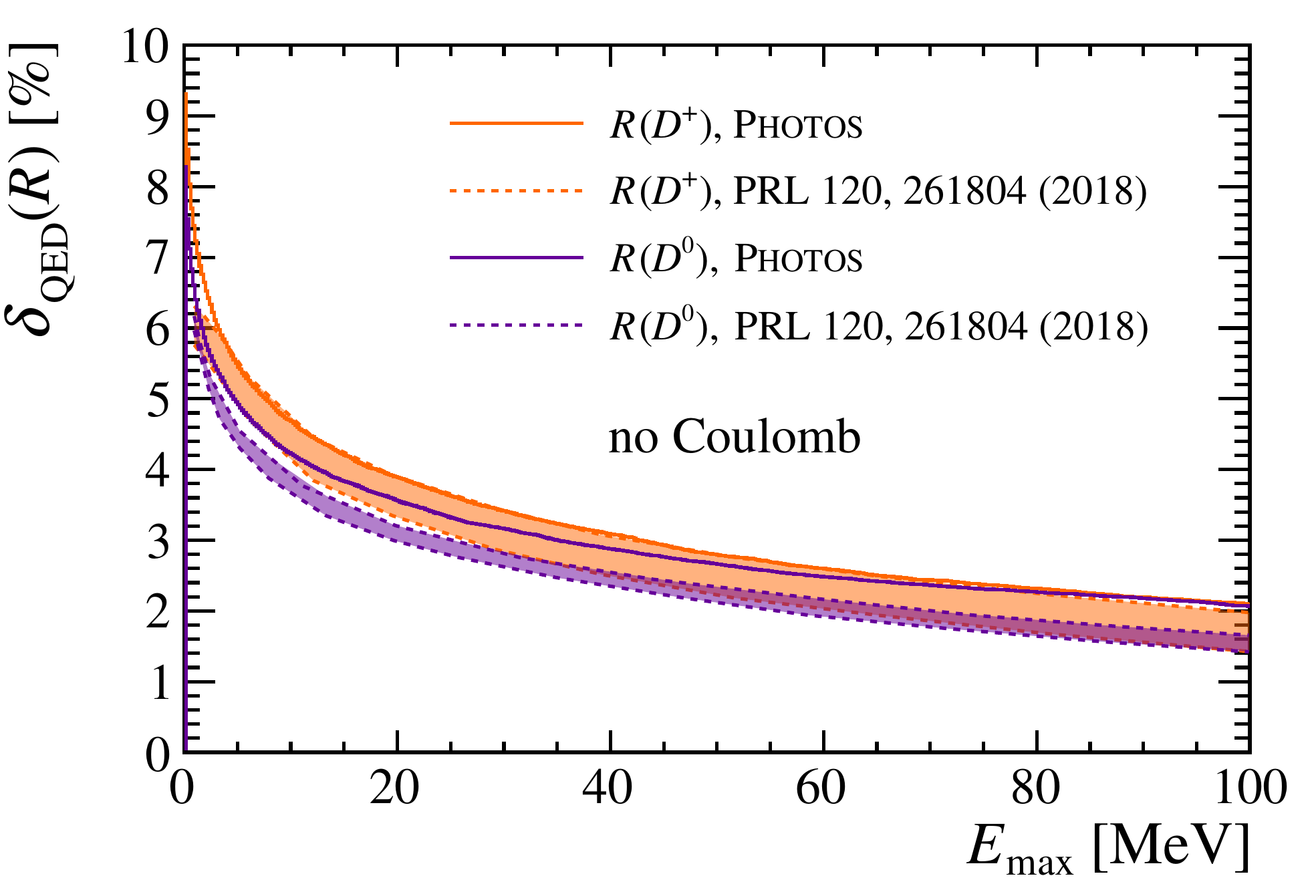}
    \includegraphics[width=0.32\linewidth]{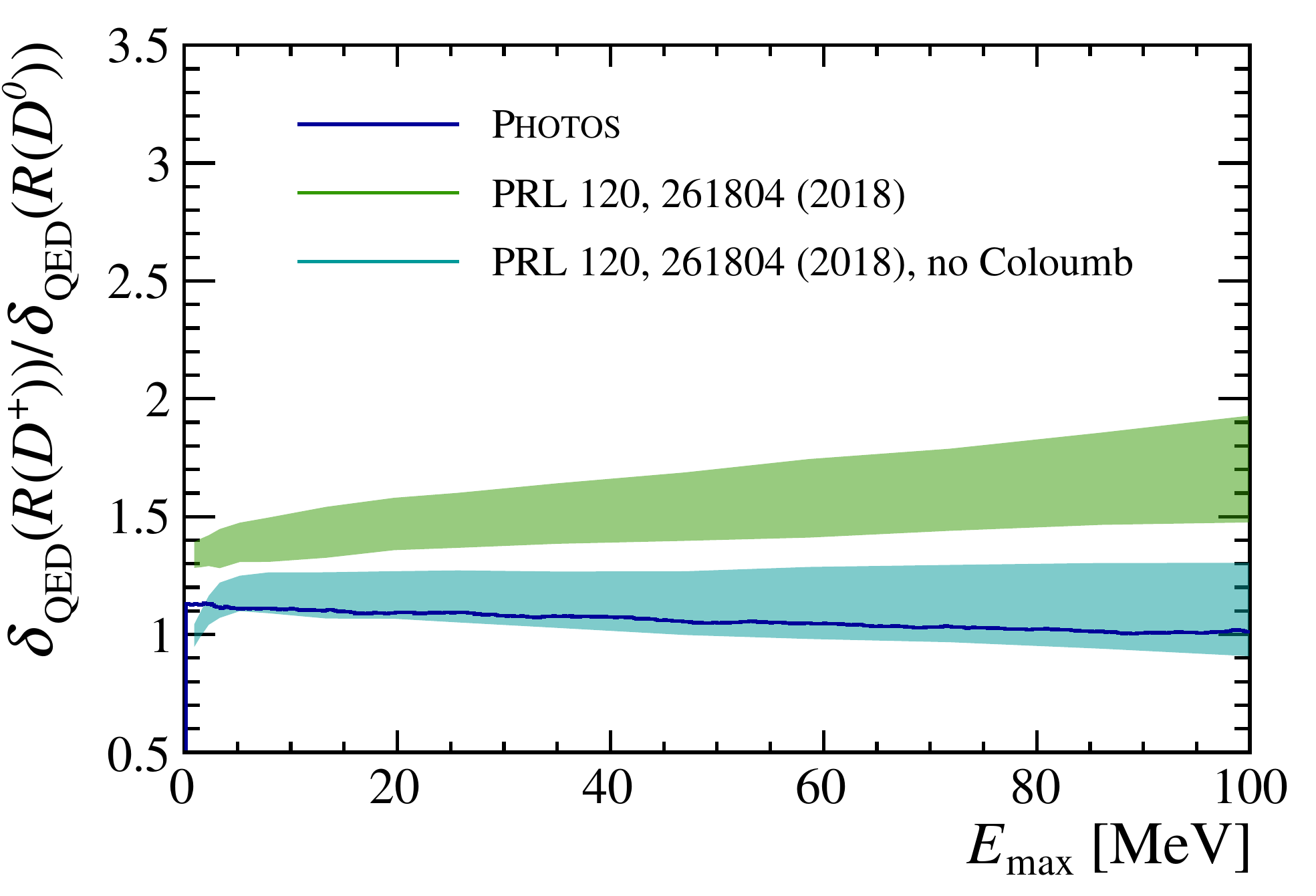}
 \vspace{-0.2cm}
  \caption{
  Radiative corrections to the branching ratios of 
  $\bar{B}^0\to D^+\ell^-\bar{\nu_\ell}$ (left) and \RDp (middle)
  in the case that no Coulomb correction is applied. The plots are 
  obtained from simulated data (solid lines) and from 
  Ref.~\cite{deBoer:2018ipi} (dashed lines). The plot on the right shows the ratio 
  $\dQED(\RDp)/\dQED(\RDz)$.}
  \label{fig:noCoulomb}
\end{figure*}

To study the differences between Ref.~\cite{deBoer:2018ipi} and \photos, four samples ($\Bzb\to\Dp\ellm\neulb$ and $\Bm\to\Dz\ellm\neulb$, 
where $\ellm = \mun, \taum$) with three million \B meson decays are generated by \pythia 8~\cite{Sjostrand:2006za,Sjostrand:2007gs}. 
The \B mesons decays are simulated by \evtgen~\cite{Lange:2001uf}, including the QED corrections by \photos v.3.56, with the ``option with interference'' switched on.
For both the $\B\to\D\mu{\bar{\nu}}_\mu$ and $\B\to\D\tau{\bar{\nu}}_\tau$ decays considered, the \verb+HQET2+ model is used with parameters from Ref.~\cite{HFLAV16}. 

The four-momentum of the total radiated photons, $\ptot_{\gamma}$, is defined as 
\begin{equation}
    \ptot_{\gamma} = \ptot_{B} - \left(\ptot_{\D} + \ptot_{\ellm} + \ptot_{\neulb} \right) \, ,
\end{equation}
where $\ptot_{B}$, $\ptot_{\D}$, $\ptot_{\neulb}$, and $\ptot_{\ellm}$ are the four-momenta of the 
\B, \D, \ellm and \neulb, respectively. 
In agreement with Ref.~\cite{deBoer:2018ipi}, the radiation of the \D decay products is not taken into account. 
The total energy of the radiated photons, $E_{\gamma}$, is computed in the \B rest frame.

The variable \emax is defined as the maximum value that $E_{\gamma}$ is allowed to have to consider $\B\to\D\ell{\bar{\nu}}_\ell(\gamma)$ signal rather than background.
The QED correction, \dQED, is given by the relative variation of the branching ratio when events with total radiated energy greater than \emax are discarded, calculated as
\begin{equation}
\dQED = \frac{\int_{0}^{\emax} N(E_{\gamma} ) dE_{\gamma} }{\int_{0}^{\infty} N(E_{\gamma}) dE_{\gamma}} - 1\, .
\end{equation}
Here, $N(E_{\gamma} )$ is the distribution of events with $E_{\gamma}$. 
The considered energy range is up to 100~MeV, covering the majority of radiative photons generated by \photos.

Figure~\ref{fig:RD0plots} shows comparisons between radiative corrections from \photos 
and Ref.~\cite{deBoer:2018ipi} to the $\Bzb\to\Dp\ellm\neulb$ (left) and $\Bm\to\Dz\ellm\neulb$ (middle) branching ratios. 
These plots show differences of up to 2\% for the \Bzb decays, and $0.5-1\%$ for \Bm decays. 
This effect does not cancel even in the ratios of branching fractions. This is clearly visible in Fig.~\ref{fig:RD0plots} (right), 
where radiative corrections on \RD, $\delta_{QED}(\mathcal{R})$, are shown as a function of \emax. 
\photos predicts a QED correction of 0.5\% lower than Ref.~\cite{deBoer:2018ipi} in \RDp, and 0.5\% higher in \RDz.

A significant part of the radiative corrections in Ref.~\cite{deBoer:2018ipi} originates
from Coulomb interactions, which are not included in \photos. 
Light leptons typically have a Coulomb correction of 1.023~\cite{Atwood:1989em},
whereas the \taum leptons in the $\Bzb\to\Dp\taum\neutb$ decay have Coulomb corrections between 2.5\% and 5.0\%.
The QED corrections from \photos for the \Dp decay mode 
are compared with predictions not including the Coulomb correction as provided by Ref.~\cite{deBoer:2018ipi}.
This reduces the difference of the corrections to the branching ratios between \photos and the theoretical calculations 
to about 1\% and brings the corrections on \RDp in close agreement, as is shown in Fig.~\ref{fig:noCoulomb}.

Figure~\ref{fig:noCoulomb} (right) shows the ratio of QED corrections on \RDp over those on \RDz. Both \photos and the calculation in Ref.~\cite{deBoer:2018ipi} without Coulomb corrections conserve isospin symmetry (\dQED values for \RDp and \RDz agree within the errors), while Coulomb corrections introduce an isospin-breaking term.
 
\section{Effects on LHCb-like analysis}
\label{sec:dummyAnalysis}

The comparison between Ref.~\cite{deBoer:2018ipi} and \photos can be made only for soft photons with energies up to 100~MeV. For higher energies, no 
calculations are available. However, \photos generates also photons with higher energies in ranges where structure-dependent photons are relevant. 
These are used to study the effects of under- or overestimating radiative corrections in simulation for a measurement of \RD in an \lhcb-like environment. 

A study is performed with the same data sets as described in the previous section by making a template fit to the 
$\B\to\D\mun\neumb$ and $\B\to\D\taum\neutb$ components. This fit uses the same fit variables as \lhcb's muonic \RDst analysis~\cite{Aaij:2015yra}.
These are the muon energy computed in the \B meson rest frame, \emu; the missing mass squared, $\mmiss = (\ptot_{\B}-\ptot_{\D}-\ptot_{\mu})^2$; and the squared four-momentum transferred to the lepton system, $\qsq=(\ptot_{\B}-\ptot_{\D})^2$. Simulated data samples are created from a mixture of $\B\to\D\mun\neumb$ and $\B\to\D\taum\neutb$ decays, with radiative corrections generated by \photos. Here \RD is assumed to be 0.3 as predicted by the SM. No backgrounds
are considered.

The fits are performed with templates that are created under the hypothesis that there is no radiation $E_\gamma$ above \emax. Five values of \emax, 
100, 300, 500, 800 and 1500~MeV, are chosen for this study. 
Fitting the templates to the simulated data sample with no cuts on radiation yields an estimate of the possible bias on \RD.
This indicates the importance of simulating $E_\gamma$ in the high-energy region. Note that \lhcb analyses do not cut on radiative energy explicitly, but that 
implicitly applied cuts could alter this bias.
 
 \begin{figure}[t]
  \includegraphics[width=0.85\linewidth]{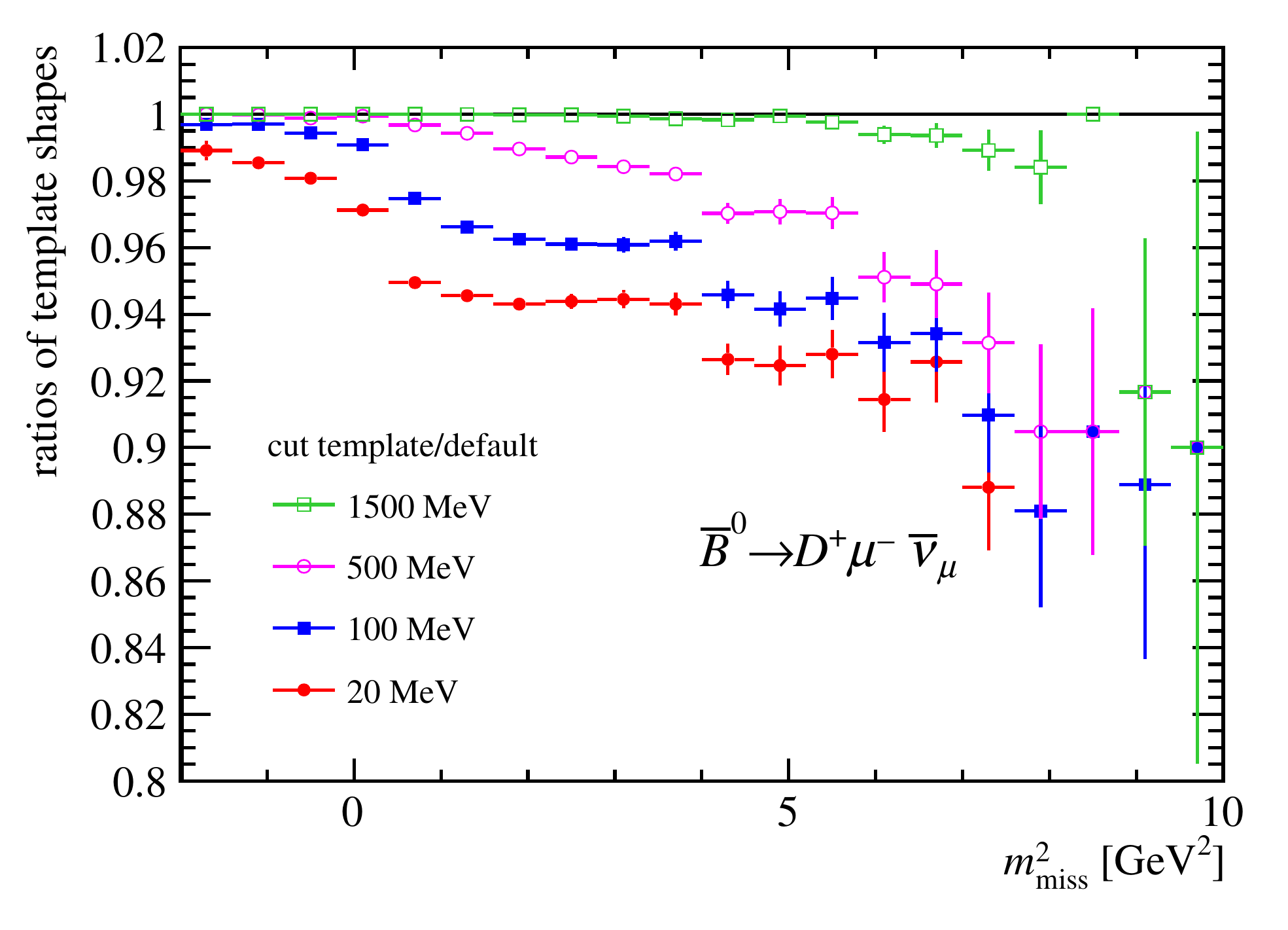}
  \vspace{-0.4cm}
  \caption{
   Ratios of the cut \mmiss distribution over the default \mmiss distribution for the  $\bar{B}^0\to D^+\ell^-\bar{\nu_\ell}$ 
  decays, for the various cuts on \emax for the \mun decay mode.}
  \label{fig:mmiss_emaxcuts}
\end{figure}

The acceptance of the \lhcb detector is mimicked by applying selection requirements following Ref.~\cite{Ciezarek:2016lqu}. 
The production and decay vertices are smeared to simulate the detector resolution and a cut on the flight distance is applied 
to resemble the trigger selection. At \lhcb, the \B meson momentum cannot be reconstructed due to the missing neutrino. 
Therefore, as in Ref.~\cite{Aaij:2015yra}, the momentum of the \B in the $z$ direction, $(\ptot_{\B})_z$, is approximated as 
$(\ptot_{\B})_z=(m_{\B}/m_{\rm vis})(\ptot_{\rm vis})_z$, where $m_{\B}$ is the \B mass,
and $m_{\rm vis}$ and $(\ptot_{\rm vis})_z$ are the mass and momentum in the 
beam direction of the visible decay products of the \B meson, respectively.

The simulated sample size is based on an estimate of the yields for the Run II data-taking period using 
the reconstruction efficiencies from Ref.~\cite{Aaij:2015yra}, the \B production cross-section at 13~TeV, 
and branching fractions. This results in yields of data samples of $1.0\times10^6$ and $0.5\times10^5$ for the 
$\Bzb\to\Dp\ellm\neulb$ decays, and $4.4\times 10^5$ and $2.3\times10^4$ for
the $\Bm\to\Dz\ellm\neulb$ decays, where the first yield represents the \mun sample, and the second the \taum sample.

\begin{figure}[h]
  \includegraphics[width=0.85\linewidth]{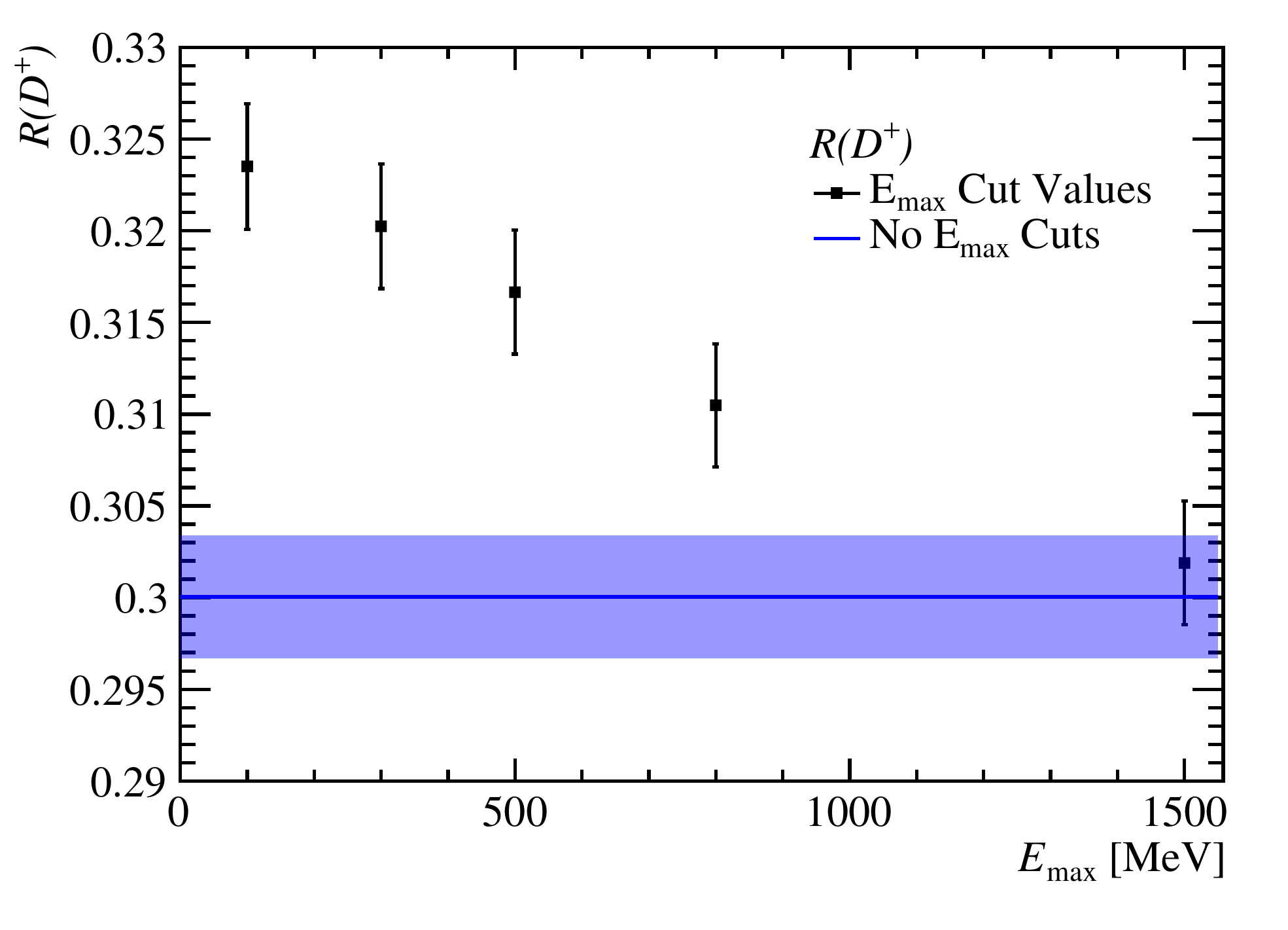}
  \vspace{-0.4cm}        
  \caption{Measured values of \RDp in a simplified \lhcb-like analysis, as a function of \emax. 
  The blue band shows the fit result obtained with the same templates used to generate the pseudo-experiments. 
  }
  \label{fig:RD_Emax}
\end{figure}

The values of \RD are determined from the fitted yields as well as the reconstruction efficiencies 
$\varepsilon_{\mu}$ and $\varepsilon_{\tau}$ for the \mun and \taum samples using 
\begin{equation}
    \RD = \frac{f_{\tau}}{1-f_{\tau}} \frac{\varepsilon_{\mu}}{\varepsilon_{\tau}} \, .
\end{equation}
It is found that for this specific case study, the ratio of efficiencies is not affected by the cuts on $E_{\gamma}$. 
Combining the efficiencies with the fitted yields, the resulting values of \RDp as a function of the cut on \emax are shown in Fig.~\ref{fig:RD_Emax}, 
which shows a dependence on \emax. From here it is clear that there is a significant effect in 
underestimating the QED radiative corrections which could be up to 0.02 for both \RDp and \RDz values, corresponding to a relative bias of 7.5\%.
This is due to the change in template shapes when applying cuts on \emax which is most clearly visible in the \mmiss distribution of the semimuonic decay, shown in 
Fig.~\ref{fig:mmiss_emaxcuts}. Since the template shapes of the semitauonic decays do not change significantly, this effect does not cancel in the ratio \RDp.

The results for \RDz analysis look very similar to those for \RDp and are therefore omitted from these proceedings. 
In actual analyses, radiative corrections are present in data, and, at least partially, in simulation. Therefore this 
analysis is likely to be an overestimate on the bias. 

\begin{figure*}[t]
    \includegraphics[width=0.32\linewidth]{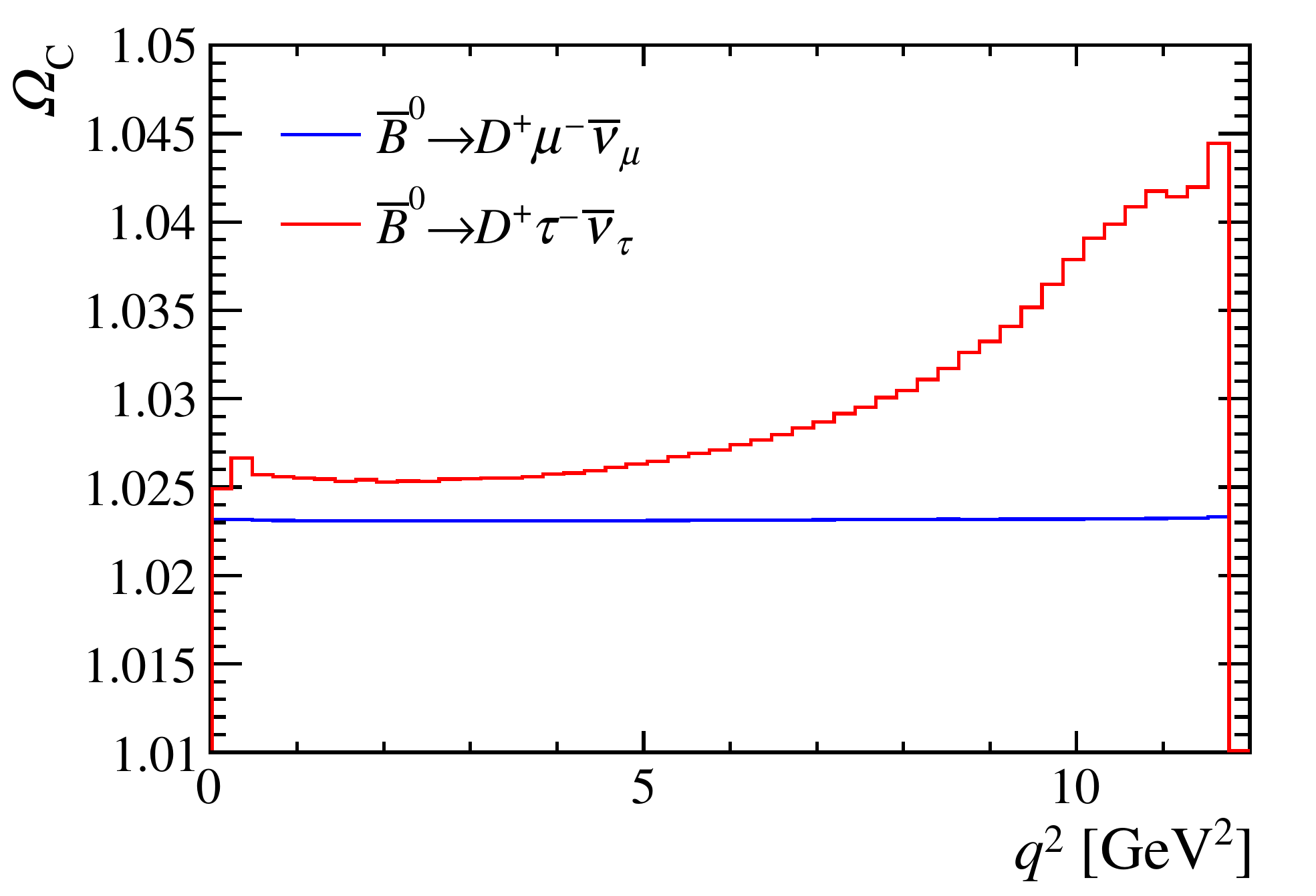}
    \includegraphics[width=0.32\linewidth]{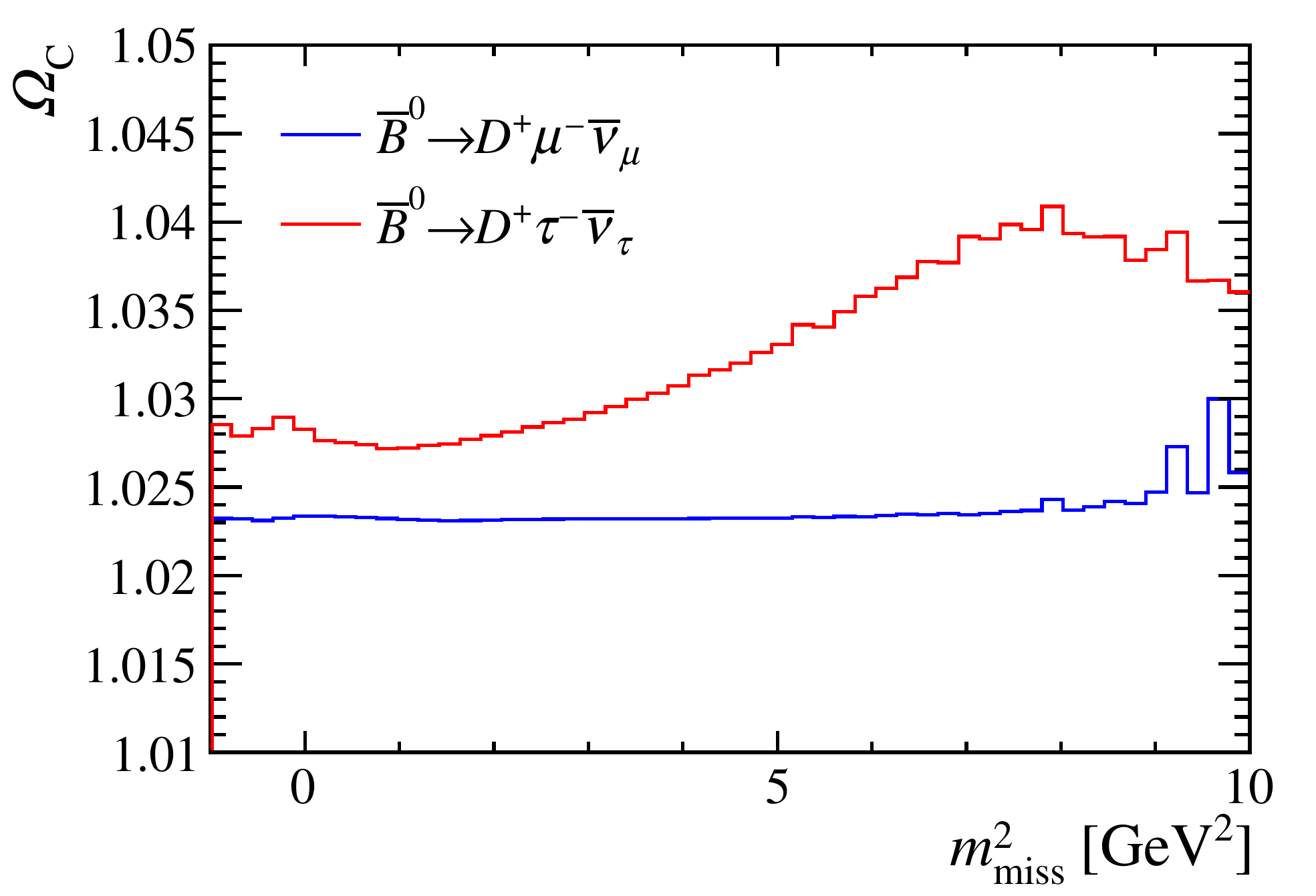}
    \includegraphics[width=0.32\linewidth]{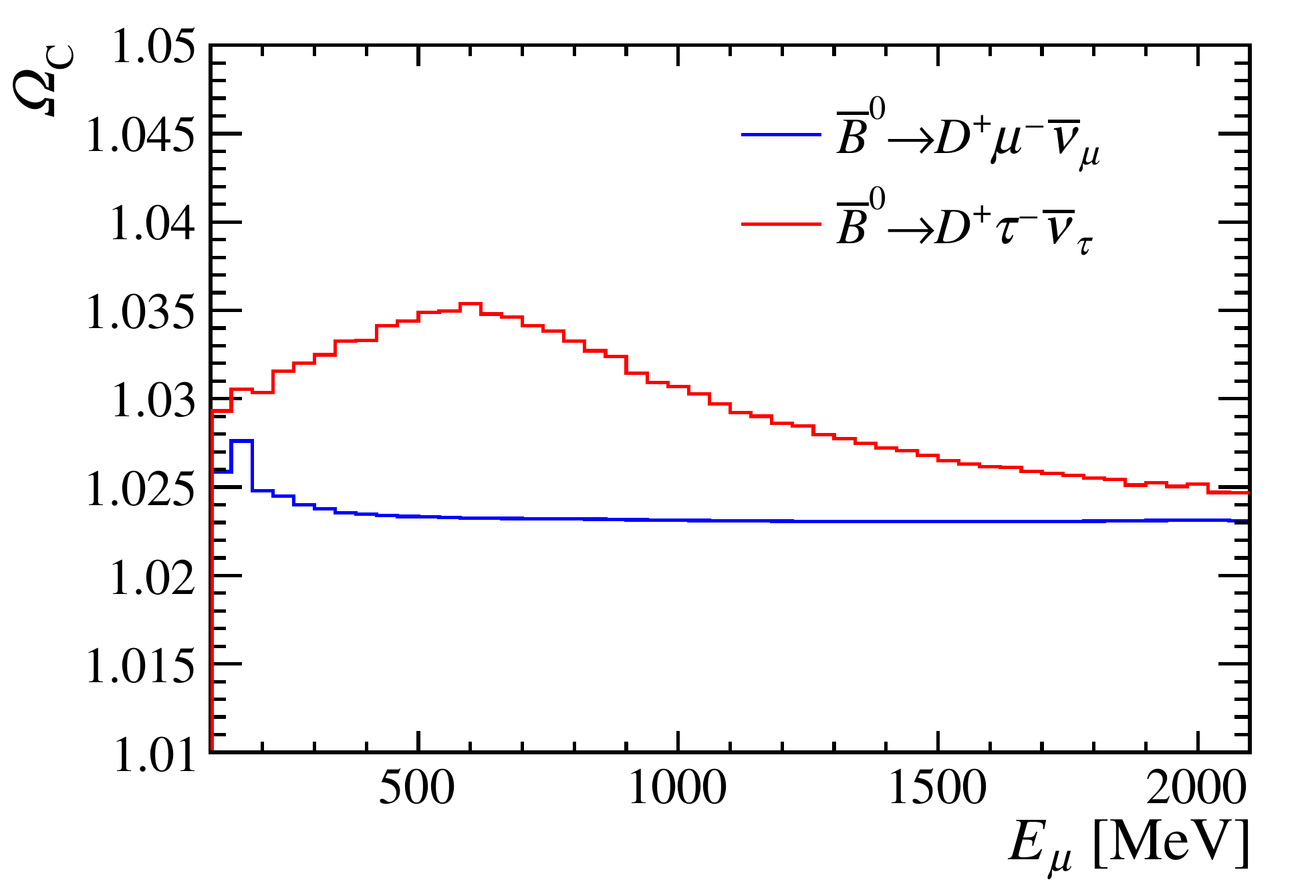}    
  \vspace{-0.2cm}
  \caption{
  Coulomb corrections as a function of \qsq, \mmiss, and \emu for the $\Bzb\to\Dp\ellm\neulb$ decays, where $\ellm = \mun, \taum$.}
  \label{fig:qsquare_Coulomb}
\end{figure*}

Also the Coulomb correction impacts the shape of the fit templates and thus the experimental results of \RDp. 
This bias is evaluated by weighting each event in the $\Bzb\to\Dp\ellm\neulb$ decay by the Coulomb correction \OmegaC.
Changes in the \qsq, \mmiss and \emu distributions are shown in Fig.~\ref{fig:qsquare_Coulomb}. \OmegaC is mostly constant for the \mun mode,
but for the \taum mode, a dependence on each of the three variables is shown. The above analysis is repeated while applying Coulomb corrections
to the simulated data sample and not on the fit templates, resulting in a relative shift on \RDp of about -1.0\%. No additional cuts on \emax are applied. 

\newpage
\section{Conclusions and recommendations}
\label{sec:Conclusions}

The QED corrections described in Ref.~\cite{deBoer:2018ipi} are not fully included in the \photos
package which is used to simulate these corrections in analyses of the \lhcb, \belle, and \babar experiments. 
These different QED corrections affect the muonic and tauonic branching ratios
at the level of a few percent. While calculating the ratios \RD, the differences largely cancel out
when Coulomb corrections are ignored. Coulomb corrections are not present in \photos and this results
in a discrepancy between Ref.~\cite{deBoer:2018ipi} and \photos of up to 1~\% on the ratio \RDp.

Coulomb interactions mainly affect the kinematics of tauonic decays, changing the shape of the 
distributions used to determine their signal yields in an \lhcb-like analysis. Not including these in simulated data can 
result in a bias of around 1\% on measurements of \RD. 
Over- or underestimating radiative corrections can bias \lhcb-like measurements up to 7\%, resulting 
in a bias of 0.02 on \RDp. 

These studies must be repeated for each analysis individually because they are
dependent on selection requirements as well as fit variables.
For \belle~II measurements, which have a better resolution on the kinematic variables~\cite{Kou:2018nap} than \lhcb, 
the effects could even be larger. 
Additional calculations of QED corrections on for $\B\to\D\ell\neul$ decays, specifically those involving high-energy and 
structure-dependent photons, are necessary in order to make measurements with higher precision. 

\begin{acknowledgments}
These proceedings are a summary of the work presented in Ref.~\cite{Cali:2019nwp} and I am very grateful to my fellow authors 
Stefano Cal\'{i}, Marcello Rotondo and Barbara Sciascia for the pleasant collaboration. 
\end{acknowledgments}

\bigskip 
\bibliography{main}

\end{document}